\documentclass[final,5p,times,twocolumn]{elsarticle}

\usepackage{graphicx}

\usepackage{amssymb}
\usepackage{afterpage}
\usepackage{wasysym}

\begin{document}

\begin{frontmatter}



\title{Large Plastic Scintillator Panels with WLS Fiber Readout; Optimization of Components}


\author[UT]{W.~Bugg}
\author[UT]{Yu.~Efremenko}
\author[UT]{S.~Vasilyev\corref{cor_author}}

\address[UT]{University of Tennessee, Department of Physics and Astronomy, 401 Nielsen Physics Building,
1408 Circle Drive, 37996-1200, Knoxville, TN,USA}

\cortext[cor_author]{Corresponding author:Phone: +1(865)438-8027;
Fax: +1(865)974-7843;
E-mail address:  svasilye@utk.edu (Sergey Vasilyev)}

\begin{abstract}
Results are presented on R\&D efforts to design and build large size veto panels, optimized for underground low background experiments,
in the most efficient and economical way using commercially available components. A variety of plastic scintillators,
photomultiplier tubes, wavelength shifting fibers, and light reflector combinations were tested. Results of these studies and performance of a 2.2 meter long panel are presented.
\end{abstract}


\end{frontmatter}



\section{Introduction}
Highly efficient and compact veto systems to tag cosmic muons become increasingly important components
for low background experiments, such as searching for Dark Matter and Double Beta decay
\cite{CDMS},\cite{PROPMAJ},\cite{MAJDEM},\cite{EXO}.
Requirements for such a system are compactness, hermiticity, very high efficiency for muon identification
and immunity from ambient gamma ray background. Often such a system has to operate in a limited
space so that large lifting fixtures to mount the system are not always an option.

Natural radioactivity present for deep underground experiments consists primarily of gamma rays from decay of
U and Th series and $^{40}$K in surrounding materials. Such gammas are normally filtered out by passive shielding
surrounding the low background detector. However, these  gammas can result in large event rates in the veto system
and therefore significant dead time if gamma and muon signals are not properly separated.
The most straightforward way to achieve good separation between gammas and muons is use of thick
scintillator \cite{KARMEN}. For scintillator with 5.10  cm  thickness a muon deposits about 10 MeV of energy which is
typically 3 times the energy of ambient gammas. Limited or non-uniform light collection can smear responses
from both gammas and muons in a veto system which is usually located outside of gamma shielding. However,  the large difference in initial energy deposition provides a suitable safety margin. This is extremely important especially for deep underground experiments where
the ratio of gamma to muon events is of the order of 10$^{7}$-10$^{9}$.

As the price of plastic scintillator (PS) is a significant factor in total detector cost it is tempting to reduce
scintillator thickness. Moving from 5.10 cm  to 2.54 cm  scintillator is a challenge because energy
depositions from muons and energetic gammas are close to each other thus increasing the chance
of misidentification. To achieve high efficiency for muons while maintaining good discrimination from gammas,
excellent uniformity and relatively large light collection is required. In addition to much lower cost, 2.54 cm
veto panels are lighter and easier to handle. For example the Majorana Demonstrator  (MJD) \cite{MNDBDE} is planning
to use large area veto panels with dimensions up to 1.3 by 1.7 m$^{2}$. The weight of such a panel built with 5.10 cm
scintillator is ${\approx}$150 kg and requires special lifting fixtures for installation. Since the final Majorana Demonstrator is housed in
a location with limited space, moving such a panel around comes a nontrivial problem.
On the other hand an identical panel built out of 2.54 cm scintillator is manageable, and can be moved
by hand.
Our R\&D effort was to establish the feasibility of constructing veto panels
using 2.54 cm scintillator while maintaining good discrimination between muons and gammas.
To avoid bulky light guides we chose to read out light from scintillators with wavelength shifting (WLS) fibers.
Systems with similar arrangements have been extensively studied primarily for accelerator based experiments \cite{MINOS},\cite{SciBar},\cite{MINERvA},\cite{T2K},\cite{GLAST}. However implementation of this technology for underground experiments raises significant
challenges, as absence of timing signals from the accelerator and/or information from multiple tracking layers complicates
the discrimination of minimum ionizing signals from ambient background.

\section{Experimental Apparatus}

\begin{figure}[!hbt]
\includegraphics[width=1.0\linewidth]{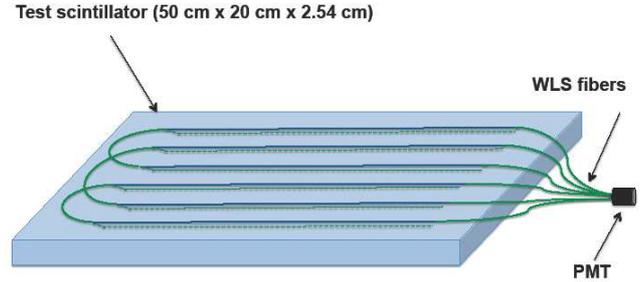}
\caption{\small Test panel}
\label{pict1}
\end{figure}

To collect light from scintillator we chose to use WLS fibers running along the scintillator in shallow grooves. An overall view
of one panel with fibers and photomultiplier tube (PMT) is shown in Fig.~\ref{pict1}. To optimize the proposed design
the R\&D program carried
out evaluation of components to select the most cost efficient combination. Five scintillators,
five wavelength shifting
fibers, two PMT models, and a variety of wrapping materials were tested. See Table~\ref{mater}.

{\renewcommand {\normalsize}{\footnotesize}
\begin{table}[!hbt]
\centering
\caption{Materials studied}
\begin{tabular}{llll}
\hline PS & PMT & WLS & REFL\\
\hline
BC-404  & R9880U-20  & Y-11(150) & TYVEK-1025D\\
BC-408  & R9880U-210 & Y-11(200) & TYVEK-1056D\\
EJ-204A &            & Y-11(300) & TYVEK HomeWrap\\
EJ-204B &            & BCF-91A   & VM 2000\\
UPS-923A    &            & BCF-92    & ESR\\
        &            &           & Aluminum Foil\\
        &            &           & Aluminized Mylar Film\\
\hline
\end{tabular}
\label{mater}
\end{table}}

BC-404 and BC-408 are plastic scintillators from SAINT-GOBAIN CRYSTALS. Important properties for BC-404 are :
light output relative to anthracene is 68\%, wavelength of maximum emission  is 408 nm and light attenuation length 140 cm. The  corresponding parameters for BC-408 are: light output 64\% ,  wavelength of maximum emission 425 nm and light attenuation length 210 cm.  EJ-204A and EJ-204B are products of
ELJEN TECHNOLOGY. They have similar properties to BC-404 but lower cost. UPS-923A is plastic scintillator from Kharkov Crystal Institute (Ukraine).
The latter scintillator has been used for the GERDA \cite{GERDA} veto system.
All test scintillator panels had dimensions 50.0 cm x 20.0 cm x 2.54 cm except UPS-923A which had thickness 3.0 cm.
The PMTs studied were HAMAMATSU photomultipliers from the R9880U series: R9880U-20 and R9880U-210 (two samples each). Both types have the same dimensions: height 12 mm, diameter 16 mm with effective photocathode diameter 8 mm.
Photocathode material for R9880U-20 is multialkali and for R9880U-210 ultra bialkali (UBA). Both are constructed with 10 dynode stages which allow typical gain 2x10$^{6}$ at 1000 V.
WLS fibers \mbox{Y-11(150)}, \mbox{Y-11(200)}, \mbox{Y-11(300)} from Kuraray and \mbox{BCF-91A},
\mbox{BCF-92} from SAINT-GOBAIN CRYSTALS were employed.
\mbox{Y-11} fiber has  diameter 1 mm and multi cladding structure.
\mbox{BCF-91A}  and \mbox{BCF-92 } fibers have diameter 1 mm and single cladding structure.
For Kuraray fibers the numbers in parentheses indicate dye
concentrations in ppm. The last column lists reflector (REFL) materials studied. TYVEK-1025D, TYVEK-1056D and TYVEK HomeWrap
are from DuPont Company and are commonly used as reflector material, VM 2000 and ESR (Enhanced Specular Reflector) are products from 3M Company and are designed to ensure high reflectivity.

The WLS fibers rested in 3 mm deep  and 1 mm wide U shaped grooves cut in the scintillator by circular saw. Only straight line grooves were cut. At the end distant from the PMT, the fibers exited the scintillator and reentered in an alternate groove. For the measurements an air gap between scintillator and WLS fibers was employed and no optical grease was used. Extensive studies were conducted to learn how to cut grooves in scintillator so that extra polishing was not required. Parameters such as saw rotation rate, scintillator feed rate and cooling water flow rates  were varied until it was possible to achieve grooves with a suitably polished surface in only  two passes of the saw. The quality of the groove surfaces was evaluated by visual inspection. The WLS fibers were looped in alternate grooves and the ends
were coupled using optical grease to a PMT secured in a special holder. See Fig.~\ref{pict1}. A light tight box was
constructed for the test measurements. Data were recorded using a LabView 8.6.1 based DAQ system. A FAST CAMAC sixteen
channel charge integrating ADC Module\- (QADC) was used to measure signal charge with integration gate 200 ns.

\section{Test of Components}
\subsection{PMT test}

\begin{figure}[!hbt]
\begin{center}
\includegraphics[width=1.0\linewidth]{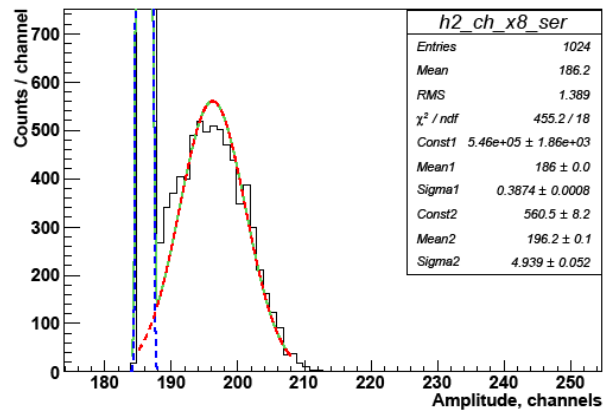}
\caption{\small Single photoelectron spectrum: dashed-blue line is fit by gaussian distribution for pedestal (parameters Const1, Mean1, Sigma1),
 and dashed-red line is fit by gaussian distribution for single photoelectron peak (parameters Const2, Mean2, Sigma2)    }
\label{pict2}
\end{center}
\end{figure}

For direct comparison of the combinations tested it is useful to express light yield in terms of number of photoelectrons.
For each test panel, gain vs. high voltage was calibrated and the position of the single photoelectron peak determined.
The single photoelectron signal was identified using a pulsed LED (blue light) with average intensity
0.01 photoelectrons per pulse. Fig.~\ref{pict2} shows the single photoelectron peak position for a typical test
panel at 920 V. The total systematic error in conversion into Phe. units is estimated to be about 15~\% based on
accuracy of single Phe. and gain measurements (see below).

\begin{figure}[!hbt]
\begin{center}
\includegraphics[width=1.0\linewidth]{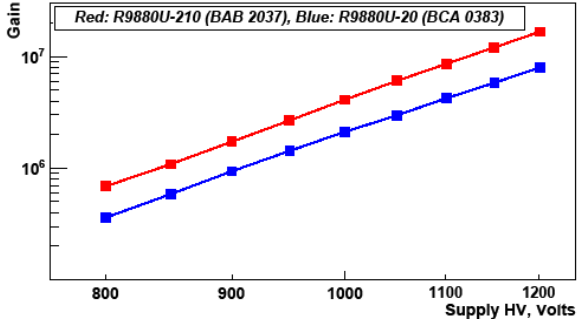}
\caption{\small Gain vs HV dependence: red line is for PMT R9880U-210 series, and blue line - for R9880U-20 series}
\label{pict3}
\end{center}
\end{figure}

For gain calibration the LED intensity was increased to about 10 Phe. per pulse as the average muon signal
is considerably larger than the single photoelectron one. An external generator was used to trigger
the DAQ and generate short LED pulses \mbox{($\le$ 50 ns)}. The results of this calibration are shown
in Fig.~\ref{pict3} for two typical PMTs, one from series R9880U-210 \mbox{(red line)} and one from series
R9880U-20 (blue line).

Two types of PMT, R9880U-20 and R9880U-210, were investigated. Measurements of muon spectra showed similar performance
with respect to gain, single Phe. spectrum, and overall quantum efficiency for WLS emission spectra.

\subsection{Measurement of attenuation in WLS fibers using muon spectrum}

\begin{figure}[!hbt]
\begin{center}
\includegraphics[width=1.0\linewidth]{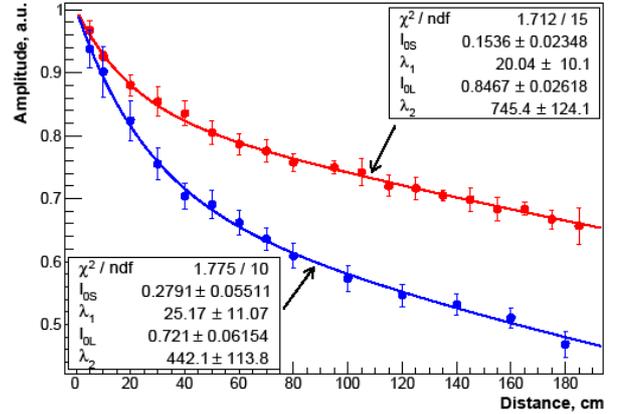}
\caption{\small Attenuation for Y-11(200) (red dots) and  BCF-91A (blue dots) WLS fibers. Attenuation length is in centimeters}
\label{pict4}
\end{center}
\end{figure}

To compare attenuation properties of different WLS fibers the cosmic ray muon peak was used.
For the measurements extra long WLS fibers were used to increase the distance between PMT and the edge of
the scintillator panel. In Fig.~\ref{pict4} the dependence of the muon peak position amplitude vs. the distance between
PMT and the edge of PS is presented for \mbox{Y-11(200)} and for  BCF-91A fibers. For the measurements BC-408, a single
\mbox{Y-11(200)}  and BCF-91A WLS fibers and REFL material TYVEK-1025D were employed. Due to the limited space in the light tight
box it was necessary to bend WLS fibers longer than 95 cm which resulted in light reduction of 2\% (measured in special runs).
Bending radius was approximately 20 cm. Correction for this effect is included in the experimental data.
Experimentally response can be fit by the sum of exponential functions.
The light output of a fiber is parameterized by:
	
	$$I(x)=I_{0S}exp(-\frac{x}{\lambda_{1}})+I_{0L}exp(-\frac{x}{\lambda_{2}})\eqno(1)$$
	
where {$\lambda_{1}$} is the short attenuation length, {$\lambda_{2}$} is the long attenuation length and $I_{0T}=I_{0S}+I_{0L}$ is the total light yield. Fig.~\ref{pict4} shows normalized experimental data for the light output of the \mbox{Y-11(200)}  (red dots) and BCF-91A (blue dots).
The attenuation data are well fit by eq. (1)  with attenuation lengths {$\lambda_{1}$} = (20$\pm$10) cm and
{$\lambda_{2}$} = (745$\pm$124) cm for \mbox{Y-11(200)} fiber and with {$\lambda_{1}$} = (25$\pm$11) cm, and {$\lambda_{2}$} = (442$\pm$113) cm for BCF-91A fiber.

Similar results have been previously reported.  For example the NOVA collaboration \cite{NOVA TDR} conducted   extensive studies of transparency of Y11 fibers vs. wavelength.   They measured attenuation length of ${\approx}$6-8 m at wave length ${\approx}$520 nm near the second maximum of emission for K27 dye (the major component of Y11 fiber). They conclude that attenuation does not depend appreciably on dye concentration but is determined by core polystyrene attenuation. We did not measure attenuation of Y11 fibers with different dye concentration.
The MINOS collaboration \cite{MINOS TDR} found that while both Kyraray (Y11) and Bicron  BCF91-A fibers were acceptable for their purposes, measurements indicated longer  attenuation length for Y11 (figure 5.10 in \cite{MINOS TDR}) than for BCF91-A.  After visual inspection of both Y11 and BCF-91A samples provided to us for tests we observe less uniformity in BCF-91A fiber.

\begin{figure}[!hbt]
\begin{center}
\includegraphics[width=1.0\linewidth]{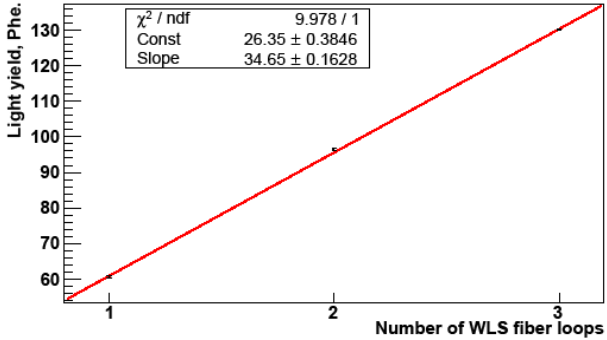}
\caption{\small Light yield as a function of the number of WLS fiber loops. Red line is linear fit to
experimental data}
\label{pict5}
\end{center}
\end{figure}

\subsection{Light yield with different numbers of WLS fibers}

Adjacent grooves in the test panel plastic scintillator are separated by 2.84 cm. Fibers are placed in one groove
and looped to return in an alternate groove. Each test panel contains 6 grooves (3 loops, see Fig. 1). The response,
when loops are added, is shown in Fig.~\ref{pict5}, where results are presented for a BC-408
panel with WLS fiber \mbox{Y-11(150)} coupled to a R9880U-20 PMT at 950 V. As expected addition of fiber loops can be
used to increase light collection.

\subsection{Light yield for various reflectors}

The dependence of light yield on wrapping materials was also investigated. For these measurements PS BC-408
with WLS fiber \mbox{Y-11(200)} coupled with PMT R9880U-20 at 950 V was used. For the three types of
TYVEK (1025D, 1056D and HomeWrap) reflectors at least two layers of material were necessary to prevent escape of light. Addition of a third layer of TYVEK increased light yield by less than 10~\%.
For reflectors VM 2000,
ESR and Aluminum Foil a single layer was used. Five layers of aluminized Mylar Film were
also tested as reflector material. As shown in Table ~\ref{Lrefl} significant improvement of light yield is observed with
VM 2000 and ESR. With these reflectors  up to \mbox{45 Phe./MeV} were collected. VM 2000 is no longer available from 3M and has been
replaced by ESR which 3M Company suggested had slightly improved reflectivity although our measurements do not confirm this for
our application.
{\renewcommand {\normalsize}{\footnotesize}
\begin{table*}[!hbt]
\centering
\caption{\small Light yield (in Phe.) at muon peak for different reflectors}
\begin{tabular}{ccccccc}
\hline Tyvek-1025D & Tyvek-1056D & Tyvek HomeWrap & VM 2000 & ESR & Aluminum Foil & Mylar Film\\
\hline\\
165.8$\pm$0.4 & 173.4$\pm$0.1 & 194. 8$\pm$0.2 & 260.3$\pm$0.3 & 246.7$\pm$0.6 & 104.7$\pm$0.3 & 64.3$\pm$0.1\\
\hline
\end{tabular}
\label{Lrefl}
\end{table*}}

\section{Results of measurements}
\subsection{Measurements on test panels}

\begin{figure}[!hbt]
\begin{center}
\includegraphics[width=1.0\linewidth]{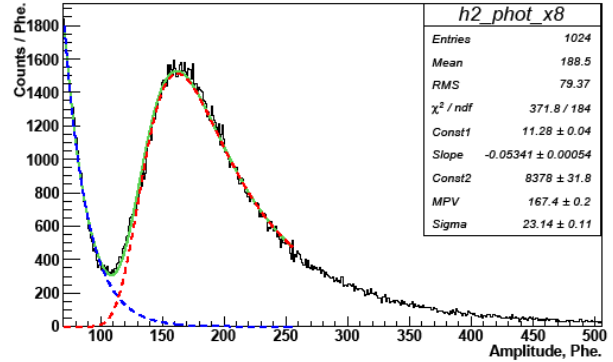}
\caption{\small Spectrum for PS BC-404. Dashed-blue line is fit by exponential distribution for environmental gammas (parameters
Const1, Slope), and dashed-red line is Landau distribution for muons (parameters Const2, MPV (most probable value), Sigma) }
\label{pict6}
\end{center}
\end{figure}

All PS-PMT-WLS combinations listed in Table ~\ref{mater} were tested. Each was wrapped individually with reflective material.
Fig.~\ref{pict6} shows a typical charge spectrum in single Phe.
units for BC-404 measured in the test laboratory at the  Science and Engineering Research Facility
(University of Tennessee, Knoxville, TN, USA). The measurement was done with WLS fiber \mbox{Y-11(200)} wrapped in
REFL material TYVEK-1025D and PMT R9880U-20 (applied voltage 950 V). The background gamma flux is shown as a
dashed-blue line and is fit to a single exponential function. It is mainly due to 2.615 MeV gamma line from $^{208}$Tl,
a product of the $^{232}$Th decay chain. The muon peak is fit by a Landau distribution, shown as the dashed-red line.

A convenient parameter for evaluation of gamma/muon discrimination is the peak to valley ratio (P/V). As can be seen
in Fig.~\ref{pict6} a larger value of this ratio represents improved separation of the gamma and muon spectra resulting
in greater muon efficiency and reduced false gamma triggers. The peak to valley ratio for the combination BC404/Y11(200), shown in Fig.~\ref{pict6}, is 4.9.

\subsection{Light yield studies}

Measurements were conducted on test panels constructed from combinations of the five PS materials and five WLS
fibers listed in Table~\ref{mater}. Each PS was wrapped with two layers of TYVEK-1025D reflector. No optical grease was used
between PS and WLS fibers. A series R9880U-20 PMT at 950 V was employed in all measurements. Light yield (in Phe.)
at the muon peak for various combinations of scintillators and fibers is shown in Table ~\ref{Lfiber}. For each
scintillator the light yield varies from worst case \mbox{(BCF-92)} to the best one \mbox{( Y-11(200))} and
\mbox{Y-11(300)}) by approximately a factor of two.

{\renewcommand {\normalsize}{\footnotesize}
\begin{table}[!hbt]
\centering
\caption{\small Light yield dependence (in Phe.) for PS WLS fiber combinations}
\setlength{\tabcolsep}{4pt}
\begin{tabular}{lccccc}
\hline PS   & Y-11(150)   & Y-11(200)   & Y-11(300)   & BCF-91A   & BCF-92\\
\hline\\
BC-404  & 121.7$\pm$0.4  & 167.4$\pm$0.2 & 160.9$\pm$0.2 & 114.9$\pm$0.2 & 78.8$\pm$0.2\\
BC-408  & 130.1$\pm$0.2  & 165.8$\pm$0.3 & 165.4$\pm$0.3 & 119.5$\pm$0.2 & 80.5$\pm$0.2\\
EJ-204A & 99.5$\pm$0.4   & 131.7$\pm$0.3 & 127.3$\pm$0.2 & 92.6$\pm$0.2  & 64.6$\pm$0.1\\
EJ-204B & 100.4$\pm$0.4  & 133.1$\pm$0.1 & 129.4$\pm$0.3 & 94.8$\pm$0.1  & 63.5$\pm$0.2\\
UPS-923A    & 54.8$\pm$0.2   & 72.2$\pm$0.2  & 74.5$\pm$0.1  & 52.3$\pm$0.1  & 34.4$\pm$0.1\\
\hline
\end{tabular}
\label{Lfiber}
\end{table}}

Table ~\ref{Lfiber} demonstrates  superior light yields  from  plastic scintillators BC-404 and BC-408
and WLS fibers \mbox{Y-11(200)}  and \mbox{Y-11(300)} .
GEANT-3 \cite{GEANT3} simulation predicts most probable energy deposition by muons in 2.54 cm scintillator to be  5.5 MeV,
indicating that the better combinations, e.g. BC404/Y11(200), provide an approximate sensitivity of 30 Phe./MeV.
To assess the reliability of the connection procedures between fibers and PMT a single BC408/Y11(200) test panel was disconnected
and reconnected 4 times. The muon peak position was measured after each step and varied less than 1~\%. The errors quoted in
Table 3 do not include possible contributions from this effect.


\subsection{P/V ratio for different PS/WLS combinations}

In Table ~\ref{PV} the P/V dependence on WLS fibers for different PS is shown.
The numerical value of the ratio, of course, depends on ambient background conditions but for a given muon to gamma ratio a large value
indicates better performance. The largest P/V ratios are for \mbox{Y-11(200)} and \mbox{Y-11(300)} WLS for nearly
all scintillators. Comparison of Table ~\ref{Lfiber} and Table ~\ref{PV} demonstrates that higher P/V values are heavily correlated with greater light yields at least for panels of the size tested. It is expected that the correlation will continue for larger panels but we have not specifically
investigated this question.

{\renewcommand {\normalsize}{\footnotesize}
\begin{table}[!hbt]
\centering
\caption{\small P/V dependence on WLS fibers for different PS}
\setlength{\tabcolsep}{4pt}
\begin{tabular}{lccccc}
\hline PS & Y-11(150) & Y-11(200) & Y-11(300) & BCF-91A & BCF-92\\
\hline\\
BC-404  & 4.5& 4.9& 4.7& 4.1& 3.2\\
BC-408  & 4.5& 5.2& 5.5& 4.5& 3.0\\
EJ-204A & 3.4& 4.3& 4.6& 3.5& 2.7\\
EJ-204B & 3.3& 4.5& 4.6& 3.6& 2.6\\
UPS-923A   & 2.7& 3.4& 3.5& 2.5& 1.8\\
\hline
\end{tabular}
\label{PV}
\end{table}}

\section{Large Panel Test}

\begin{figure}[!hbt]
\begin{center}
\includegraphics[width=1.0\linewidth]{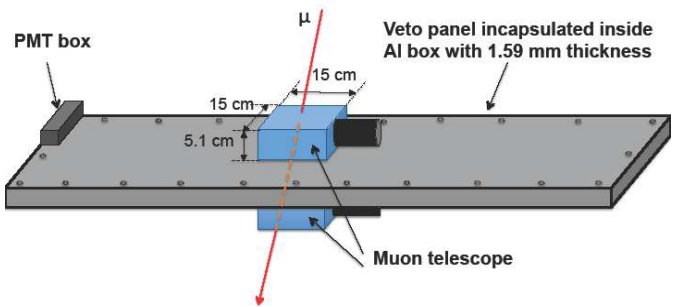}
\caption{\small Veto panel 224 cm x 33 cm x 2.54 cm in Aluminum box with muon telescope}
\label{pict7}
\end{center}
\end{figure}

The focus of our R\&D was to determine optimal technology for construction of large veto panels for the Majorana Demonstrator ~\cite{MNDBDE}.  An important requirement is reduction of dead areas to an absolute minimum to have a good geometrical efficiency for muons. To achieve this goal the PMT is located above the plane of scintillator.
While a variety of panel lateral dimensions are required in the Demonstrator, a typical panel is 170 cm long, 80 cm wide with
thickness 2.54 cm. To explore the properties of larger panels we constructed a panel with the dimensions of the longest panel required by the Majorana Demonstrator, which is shown in Fig.~\ref{pict7}. Five fibers extend the full length of the panel and return
through an alternate groove. The total number of grooves is ten. Fibers in the panel
are bunched together at one end, bent upward with 6 cm radius  and connected to the  PMT. The scintillator plate is wrapped in reflective material
and enclosed in a light tight Aluminum box (Al thickness=1.59 mm). A single PMT was used in contrast to the widely employed system with PMTs attached to the both ends of the fibers.  This reduces cost of electronics, PMTs, cables, etc. and ensures less dead space around veto panels. Wavelength shifting fibers ride freely in the
scintillator grooves without optical coupling.
Uniformity of response to muons was measured using a small movable
scintillator telescope shown in Fig.~\ref{pict7} to select cosmic muons passing through the panel at specific locations.

\begin{figure}[!hbt]
\begin{center}
\includegraphics[width=1.0\linewidth]{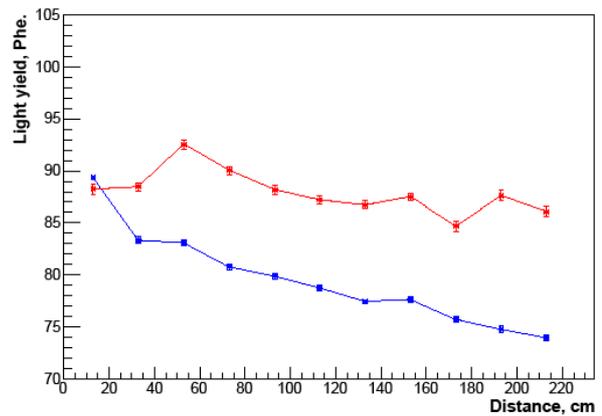}
\caption{\small Panel uniformity light collection. Blue line is a uniformly wrapped panel and red line is with optimized two component wrapping}
\label{pict8}
\end{center}
\end{figure}

Fig.~\ref{pict8} shows uniformity of light collection from the 224 cm long panel. Uniformity was measured by scanning along the
panel with the muon telescope in steps of 20 cm. The vertical scale is the most probable light output for muons in photo-electrons. The horizontal scale is distance from the end where PMT is located. Blue points are response of
the panel with scintillator wrapped uniformly with two layers of TYVEK-1025D. The fact that wrapping materials have different light collection efficiencies can be used to improve panel response uniformity. Use of highly reflective material at the far end of panels with less reflective material near the  PMT can significantly improve light collection uniformity. The effect of such optimized wrapping is shown in Fig.~\ref{pict8}.
Red points are response of the panel with scintillator wrapped in a combination of TYVEK-1025D and ESR reflectors optimized to improve
uniformity. This non-uniform wrapping improved response uniformity (2.2~\% instead of 5.6~\%) and average light output increased. We
have consistently noted for all of the larger panels constructed for MJD that light output is reduced from that observed in the small test
panels. We are unable to fully account for this phenomenon but several factors seem to contribute, including wider spacing of grooves and longer WLS fiber lengths needed to reach the PMT.

\begin{figure}[!hbt]
\begin{center}
\includegraphics[width=1.0\linewidth]{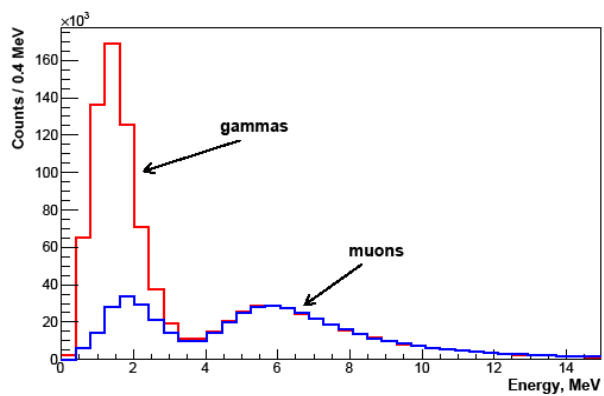}
\caption{\small Self triggering spectra by low threshold (red line) and high threshold (blue line)}
\label{pict9}
\end{center}
\end{figure}

In Fig.~\ref{pict9} the result of a run in self triggering mode from the same panel is shown for two thresholds.
Low \mbox{(red line)} threshold was chosen to detect all muons and consequently many ambient gammas. Clear separation is seen between muons and gammas. A higher threshold (blue line) retains 99~\% of muons but suppresses the gamma rate
by factor of 4. The trigger rate for gammas is then less than 10 Hz/panel which introduces negligible dead time into the system.
For trigger threshold we used an amplitude discriminator while light yield was determined by total charge. Due to fast response
of the PMT and the relatively large spread in arrival time of photons in a large scintillator, a low amplitude signal can correspond
to a large total energy deposition. Therefore in Fig.~\ref{pict9} we do not observe a sharp energy cut.


\section{Conclusions and discussions}

A comprehensive study of combinations of various components used to build large veto panels is presented.
It is demonstrated that with 2.54 cm  scintillator and a single photomultiplier it is possible to achieve excellent
light collection uniformity, large light output and good gamma to muon separation even for scintillation panels more
than 2 m long. The best performance of panels for maximum light output and P/V ratio is
the combination of \mbox{(BC-408)+(Y-11(200))+(VM 2000)}. However, it is also the
 most expensive, and due to budgetary constrains we were not able to select this as a baseline for MJD.
 From all components the most expensive is scintillator which would account for 80~\% of the cost if BC-408 is selected. Since the prices quoted for the scintillators tested varied by nearly a factor of four, the scintillator choice was crucial and resulted in selection of EJ-204B even though the muon light yield was lower by 25-30~\%. Based on these considerations we selected combination of  EJ-204B scintillator, Y-11(200) WLS, optimized wrapping
 \mbox{(ESR+TYVEK)}, and R9880U-210 PMT as a baseline for MJD. With this choice the use of 2.54 cm thick scintillator relative to 5.10 cm reduced the final cost of the MJD veto system by 60~\%.

{\vspace{4mm}\bf Acknowledgments \vspace{4mm}}

This work was supported by DOE grant number DE-FG02-10ER41715
 (USA). Authors are thankful to members of the Majorana Collaboration for support and interest in those studies.








\end{document}